\documentclass[aps, preprint,groupedaddress,showpacs]{revtex4}
\usepackage{color}
\usepackage{graphics}
\usepackage{natbib}
\begin{document}
%%\renewcommand{\familydefault}{cmss}
%%\sffamily
\newcommand{\beq}{\begin{equation}}
\newcommand{\eeq}{\end{equation}}
% You should use BibTeX and revtex.bst for references
\bibliographystyle{apsrev}

\title{Preferred states of the apparatus}

%\homepage[]{Your web page}
\author{Anu Venugopalan}
\email{anu.venugopalan@gmail.com}
\affiliation{School of Basic and Applied Sciences, GGS Indraprastha University, \\ Sector 16C, Dwarka, New Delhi-110 075, India}

%%\date{\today}

\begin{abstract}

A simple one dimensional model for the system-apparatus interaction is analyzed. The system is a spin-1/2 particle, and its position and momentum degrees constitutes the apparatus. An analysis involving only unitary Schr\"{o}dinger dynamics illustrates the nature of the correlations established in the system-apparatus entangled state. It is shown that even in the absence of any environment-induced decoherence, or any other measurement model, certain initial states of the apparatus - like localized Gaussian wavepackets - are preferred over others, in terms of the establishment of measurementlike one-to-one correlations in the pure system-apparatus entangled state. 

\vspace{0.5cm}

{\bf Keywords} 

\vspace{0.5 cm}

{\bf Quantum measurement, system-apparatus correlations, pointer states, preferred basis}
\end{abstract}

\pacs{03.65.Ud ; 03.65.Ta}

\maketitle

\section{Introduction}
Though quantum theory is widely accepted as the fundamental theory of nature and is immensely successful and satisfying, its conceptual framework makes predictions which are difficult to comprehend ``classically". Many of these conceptual problems are encompassed in what is known as the quantum measurement problem\cite{dirac, bohm, peres, wz}. The quantum mechanical description of a system is contained in its wave function or state vector $|\psi\rangle$ which lives  in an abstract ``Hilbert space". The dynamics of the wavefunction is governed  by the Schr\"{o}dinger equation
\begin{equation}
i\hbar \frac{d}{dt}|\psi\rangle=H|\psi\rangle.
\end{equation}
Here $H$ is the Hamiltonian of the system and the equation is linear, deterministic and the time evolution governed by it is unitary.  Dynamic variables or {\it observables} are represented in quantum mechanics by {\em linear Hermitian operators} which act on the state vector. An operator $\hat{A}$, corresponding to the dynamical quantity $A$ is associated with {\em eigenvalues} $a_{i}$ and corresponding {\em eigenvectors} $\{\alpha_{i}s\}$ which form a {\em complete orthonormal set}. Any arbitrary state vector, $|\psi\rangle$ can, in general, be represented as a linear superposition of these eigenvectors:
\begin{equation}
|\psi\rangle=\Sigma c_{i}|\alpha_{i}\rangle.
\end{equation}
A basic postulate of quantum mechanics regarding {\em measurement} is that any measurement of the quantity A can only yield {\em one} of the eigenvalues, $a_{i}$s,  but the result is not definite in the sense that different measurements for the quantum state  $|\psi\rangle$ can yield different eigenvalues. Quantum theory predicts only that the {\em probability} of obtaining eigenvalue $a_{i}$ is $|c_{i}|^{2}$.
The {\em expectation value} of the operator $\hat{A}$ is defined in quantum mechanics as:  
\begin{equation}
\langle\hat{A}\rangle=\langle\psi|\hat{A}|\psi\rangle=\Sigma{ a_{i}|c_{i}|^{2}}.
\end{equation}
In terms of the density matrix $\hat{\rho}=|\psi\rangle\langle\psi|$, an equivalent formula for the expectation value is:
\begin{equation}
\langle\hat{A}\rangle=Trace \{\hat{A} \hat{\rho} \}.
\end{equation}
An additional postulate of quantum mechanics is that the measurement of an observable A, which yields one of the eigenvalues $a_{i}$ ( with probability $|c_{i}|^{2}$) culminates with the {\em reduction } or {\em collapse } of the state vector $|\psi \rangle$ to the eigenstate $| \alpha_{i}\rangle$. This means that every term in the linear superposition vanishes, except one. This reduction is a {\em non unitary process} and hence in complete contrast to the unitary dynamics of quantum mechanics predicted by the Schr\"{o}dinger equation and this is where the crux of the conceptual difficulties encountered in quantum theory lies. These two stages of quantum measurement are captured in the well-know von Neumann model through two distinct processes:(1) the system-apparatus interaction via linear unitary Schrodinger evolution via an appropriate interaction Hamiltonian and (2) the nonlinear, indeterministic collapse \cite{von}.
\noindent
A quantum mechanical description for a measurement process would typically involve a coupling between the microscopic system and a 'macroscopic' apparatus (meter), resulting in an {\em entangled state}  - a uniquely quantum mechanical state for the composite. This entangled state should contain one-to-one correlations between the states of the system, $\{|\psi_{Si}\rangle\}$, and the states of the apparatus $\{|\phi_{Ai}\rangle\}$, so that a read out of the apparatus or `meter' states gives information about the states of the system. For example, for a simple two-level system, an entangled system-apparatus state should look like
\begin{equation}
|\Phi_{S+A}\rangle=|\psi_{S1}\rangle|\phi_{A1}\rangle + |\psi_{S2}\rangle|\phi_{A2}\rangle.
\end{equation}
Of course, the problem with such an entangled state is that it seems to allow the 'meter' to exist in a coherent superposition of states which could be macroscopically distinct - a situation hard to reconcile with classical intuition. The density matrix corresponding to this entangled sates is;
\begin{eqnarray}
\hat{\rho}_{S+A}&=&|\psi_{S1}\rangle\langle\psi_{S1}||\psi_{A1}\rangle\langle\psi_{A1}|+|\psi_{S2}\rangle\langle\psi_{S2}||\psi_{A2}\rangle\langle\psi_{A2}|\nonumber \\  +&&|\psi_{S1}\rangle\langle\psi_{S2}||\psi_{A1}\rangle\langle\psi_{A2}|  + |\psi_{S2}\rangle\langle\psi_{S1}||\psi_{A2}\rangle\langle\psi_{A1}|.\nonumber \\ 
\end{eqnarray}
While (6) represents a perfectly legitimate solution of the Schr\"{o}dinger equation, the physical interpretation in the usual language of probabilities leads to difficulties. While the diagonal elements ( the first and second terms) can be easily interpreted as probabilities corresponding to the system being in state $|\psi_{S1}\rangle$ or $|\psi_{S2}\rangle$  (with the corresponding correlations with the apparatus states $|\phi_{A1}\rangle$ and $|\phi_{A2}\rangle$, respectively), the off-diagonal elements represented by the third and fourth terms are difficult to interpret classically in terms of probabilities. In order to make 'classical' sense, the density matrix corresponding to the pure state ensemble described by this entangled state must reduce to a {\em statistical mixture} which is diagonal in some basis with appropriate system-apparatus correlations. Several interpretations of quantum mechanics seek to explain this $\rho_{pure} \rightarrow \rho_{mixed}$ transition and a resolution to the mechanism for the apparently nonunitary 'collapse' in a quantum measurement\cite{wz,zurek1}. In recent years, the decoherence approach \cite{joos} has been widely discussed and accepted as the mechanism responsible for this transition. The central idea of this approach has been that 'classicality' is an emergent property of systems interacting with an environment. The theory also predicts that in a quantum measurement, the apparatus will have correlations with the system in a set of 'preferred states'\cite{zurek1, zurek2, joos} selected by the environment. (It is worth mentioning here that a criticism often levelled against the decoherence theory is that even though it allows the transition from a {\em pure} to a {\em mixed} state, it does not explain the fact that only one of the mixed outcomes is actually observed.)

In the following, we will not discuss the various approches to resolving the measurement problem, but focus, instead on just the simple situation of a system-apparatus entangled state formed during a measurement-like interaction under pure unitary evolution. Our study shows that even in the absence of any environment-induced decoherence, or any other measurement model, certain initial states of the apparatus - like localized Gaussian wavepackets - are preferred over others, in terms of the establishment of measurementlike one-to-one correlations and the reduction in magnitude of the off-diagonal elements (even though they would never `vanish' as the dynamics is completely unitary). This simple model is discussed and analyzed in the next section.

\section{A simple model for measurement}
Consider a simple model of measurement consisting of a free particle with spin (for simplicity, consider a two-state system or spin-$1/2$ which could represent a qubit). Here the spin degrees of freedom represent the system and the position and momentum degrees represent the apparatus.The system and apparatus are coupled by a Stern-Gerlach measurementlike interaction such that the trajectory of the particle (position and momentum degrees) correlates with the spin states. The Hamiltonian describing this model is:
\begin{equation}
H=\lambda \sigma_{z} + \frac{p^{2}}{2m} + \epsilon z \sigma_{z}.
\end{equation}
While the first two terms represent the self Hamiltonians of the system and apparatus, respectively, the last term is the interaction Hamiltonian. $z$ and $p$ denote the position and momentum of the particle of mass $m$, $\lambda \sigma_{z}$ the Hamiltonian of the system and $\epsilon$ the product of the field gradient and the magnetic moment of the particle. The most general initial state for the system-apparatus combine can be written as
\begin{equation}
\psi= \{a|\uparrow\rangle + b|\downarrow\rangle\} \otimes \phi(z).
\end{equation}
This is a product state of the most general spin state for the system and an arbitrary state, $\phi(z)$ for the apparatus (free particle). $|\uparrow\rangle$ and $|\downarrow\rangle$ are the eigenstates of $\sigma_z$. Following the measurement interaction governed by  Hamiltonian evolution, this initial state becomes an {\it entangled state} between the system and apparatus, whose density matrix can be written as:
\begin{eqnarray}
\rho_{S+A}&=& |a|^{2}|\uparrow\rangle\langle \uparrow| \rho_{\uparrow \uparrow}(z, z', t) + |b|^{2}|\downarrow\rangle\langle \downarrow| \rho_{\downarrow \downarrow}(z, z', t) \nonumber \\ +&&  ab^{*} |\uparrow\rangle\langle \downarrow| \rho_{\uparrow \downarrow}(z, z', t) + a^{*}b|\downarrow\rangle\langle \uparrow| \rho_{\downarrow \uparrow}(z, z', t).
\end{eqnarray}
Here $\rho_{\uparrow \uparrow}$ and $\rho_{\downarrow \downarrow}$ correspond to  the diagonal elements (in spin)  of the density matrix ($\rho_{d}$) for the apparatus which could correlate with up and down spin states of the system and $\rho_{\uparrow \downarrow}$ and $\rho_{\downarrow \uparrow}$ correspond to the off-diagonal elements ($\rho_{od}$), and $\rho(z, z', t) = \langle z |\rho|z'\rangle$.

The specific form of $\rho _{d}$ and $\rho_{od}$ and the system correlations they would (or would not) contain depends on the initial state of the apparatus. When they contain one-to-one system-apparatus correlations, the states corresponding to $\rho_d$ would be candidate {\em pointer states}. Pointer states have been widely discussed in detail in the context of decoherence \cite{zurek3, av}.  Here we use the term only to refer to those states which will contain correlations with the system states, and hence have the potential to effect a {\em measurement}. In spite of the simplicity of the model considered here, it is not obvious what these states would be and whether any arbitrary initial state of the apparatus will lead to a system-apparatus entangled state with one-to-one measurementlike correlations. In the context of environment induced decoherence, it has been suggested that when the intrinsic dynamics of the system and apparatus can be neglected, pointer states would turn out to be the eigenstates of the interaction Hamiltonian of the apparatus-environment (or simply system-environment, in the absence of the apparatus)\cite{zurek3}. In many such models, the interaction Hamiltonian is diagonal in the position basis and hence position is expected to be the pointer basis. However, it turns out that this is not necessarily so, as many studies show\cite{zurek3,av}. For the simple measurement-like scenario described by (7) which does not involve any coupling to the environment, the pointer basis can be understood as those states of the apparatus in which correlations with the system states are eventually established. This is  a necessary step before environmental decoherence reduces the post-measurement density matrix of the composite system to its appropriate diagonal form and hence the nature of these correlations are very important. It is intuitive that the pointer states should naturally be a consequence of the interplay between the various components of the total Hamiltonian. If the measurement time scales are small compared to the intrinsic time scales of the system, it is physically reasonable to ignore the  self Hamiltonian of the apparatus in (7). This corresponds to the trivial situation where the evolution is governed simply by
\begin{equation}
H=\epsilon \sigma_{z} z.
\end{equation}
In the more general situation, where the intrinsic dynamics of the system is relevant, an interplay between the three terms in (7) should give rise to the pointer basis. It is not intuitively obvious what these would be. The general solutions for unitary evolution via the Hamiltonian in Eq. (7) for $\rho_{d}$ and $\rho_{od}$ in the partial Fourier transform representation are:
\begin{equation}
\rho_{d}(Q, r, t)=\rho(Q, r+ \frac{\hbar Qt}{m},0) \exp{\big (\mp \frac{i\epsilon rt}{\hbar}\mp \frac{i \epsilon Q t^{2}}{2m}\big)},
\end{equation}
\begin{equation}
\rho_{od}(Q, r, t)=\exp{\pm\frac{2i\lambda t}{\hbar}}\rho(Q \pm \frac{2\epsilon t}{\hbar}, r + \frac{\hbar Q t}{m} \pm \frac{\epsilon t^{2}}{m}, 0), 
\end{equation}
for the case where the self Hamiltonian is included, and
\begin{equation}
\rho_{d}(Q, r, t)=\rho(Q, r,0) \exp {\pm\frac{i\epsilon r t}{\hbar}},
\end{equation}
\begin{equation}
\rho_{od}(Q, r, t)=\rho(Q \pm\frac{2 \epsilon t}{\hbar}, r,0),
\end{equation}
when the self Hamiltonian is ignored. Here, the new variables are
\begin{eqnarray}
R=\frac{z+z'}{2}, \nonumber \\
r=z-z', \nonumber \\
Q=p-p', \nonumber \\
q=\frac{p+p'}{2},\nonumber \\
\rho(Q, r, 0)=\int_{-\infty}^{\infty}e^{iQR}\rho(R, r,0)dR,
\end{eqnarray}
where $z$ refers to the position coordinate and $p$ to the momentum. Note that while $\rho(z,z')=\langle z|\rho|z'\rangle$, $\rho(R,r)$ is $\rho(z,z')$ expressed in terms of the new variable $R$ and $r$ and similarly, $\rho(Q,q)$ is $\rho(p,p')$ expressed in terms of the new variables $Q$ and $q$.
In the next section, these solutions are  analysed as applied to three specific initial states of the apparatus: a momentum eigenstate, a position eigenstate and a Gaussian wavepacket. This simple analysis illustrates the type of measurementlike correlations that could develop between the system and the apparatus states and sheds light on the nature of the pointer basis that could emerge.

\section{Analysis}
\subsection{Apparatus in initial momentum eigenstate}
One starts off with an initial condition where the apparatus (free particle) is in a momentum eigenstate
\begin{equation}
\psi=e^{ikz}.
\end{equation}
The  initial density matrix in the partial Fourier transform representation corresponding to Eq. (16) will be:
\begin{equation}
\rho(Q, r, 0)=e^{ikr} \delta(Q).
\end{equation} 
Note that this state is not normalizable and hence not the ideal choice for a real apparatus. However, it is still useful to explore the nature of the correlations that are developed between the system and apparatus for such an initial state. It is easy to see that the solutions for the diagonal and off-diagonal elements (in spin) of the density matrix (in the partial FT representation) are:
\begin{eqnarray}
\rho_{d}(Q, r, t)& =& \exp{\{ik (r + \frac{\hbar Q t}{m})\mp\frac{i \epsilon rt}{\hbar} \mp \frac{i\epsilon Q t^{2}}{2m}\}} \delta(Q),  \\
\rho_{od}(Q, r, t)&=& \exp{\{\pm \frac{2 i \lambda t}{\hbar} + ik (r + \frac{\hbar Q t}{m} \pm \frac{i \epsilon t^{2}}{m}) \}} \nonumber \\
&&\otimes \quad \delta (Q \pm \frac{2 \epsilon t}{m}), 
\end{eqnarray}
The solutions for the diagonal elements in position and momentum representation are:
\begin{eqnarray}
\rho_{d}(R,r,t)&=& \exp {\{(ik \mp\frac{\epsilon t}{\hbar})r\}},  \\
\rho_{d}(Q, q,t)&=& \delta(q + k \mp \frac{\epsilon t}{\hbar})\delta(Q) \nonumber \\
&& \otimes  \exp{\{\frac{i\hbar k Q t}{m} \mp \frac{i\epsilon Q t^{2}}{2m}\}}.\end{eqnarray}
Equations (20) \& (21) represent the apparatus (pointer) states in position and momentum representations which should be correlated  with the up and down spin states of the system to effect a measurement. If one looks at the diagonal elements in position, i.e., setting $R=z, r=0$, then it is clear from (20) there are {\it no one-to-one correlations} between the system and the apparatus states. (20) contains no signatures of the 'up' or 'down' states of the spin and hence can record no measurement. In the momentum representation, (21), the diagonal elements are obtained by setting $Q=0, q=p$, and this gives the pointer states, $\rho_{\uparrow \uparrow},\rho_{\downarrow \downarrow}$ as $\delta (p+k \mp \frac{\epsilon t}{\hbar}$). The 'up' and 'down' spin states of the system correlate with the apparatus states through the $\mp \frac{\epsilon t}{\hbar}$ terms which can be interpreted as momentum kicks:
\begin{eqnarray}
\rho_{S+A}&=& |a|^{2}|\uparrow\rangle\langle \uparrow|\delta (p+k - \frac{\epsilon t}{\hbar})  + \nonumber \\
&& |b|^{2}|\downarrow\rangle\langle \downarrow| \delta (p+k + \frac{\epsilon t}{\hbar}) \nonumber \\ +&&  ab^{*} |\uparrow\rangle\langle \downarrow| \rho_{\uparrow \downarrow} + a^{*}b|\downarrow\rangle\langle \uparrow| \rho_{\downarrow \uparrow}.
\end{eqnarray}
The correlations here are clearly in the momentum basis. The off-diagonal elements of the density matrix in the partial FT representation are:
\begin{equation}
\rho_{od}=e^{\frac{2i\lambda t}{\hbar}} \delta (Q \pm \frac{2 \epsilon t}{m}) \exp{\{r + \frac{\hbar Q t}{m} \pm \frac{i \epsilon t^{2}}{\hbar}\}}.
\end{equation}
If the time scale for the measurement is extremely small compared to the intrinsic time scales in the system the self Hamiltonian can be ignored and the solutions for $\rho_d$ with just the interaction Hamiltonian ( last term in Eqn. (7)) in position and momentum representations are:
\begin{eqnarray}
\rho_{d}(R,r,t)&=& \exp {\{(ik \mp\frac{\epsilon t}{\hbar})r\}},  \\
\rho_{d}(Q, q,t)&=& \delta(q + k \mp \frac{\epsilon t}{\hbar})\delta(Q).
\end{eqnarray}
A look at the diagonal elements show that the pointer states are exactly the same as when the self Hamiltonian is included.  It may be noted that this scenario involving only the interaction Hamiltonian with the initial state of the apparatus in the momentum eigenstate is that envisaged in the first stage of the well known von Neumann measurement scheme\cite{von}. Infact, in this scheme one can see tht since the interaction Hamiltonian in (7) is diagonal in position, the pointer states will indeed be in the conjugate basis, i.e., momentum.

It is interesting to note that the pointer states are exactly the same with and without the self Hamiltonian. Unlike the requirement of short interaction time in the case when only the interaction Hamiltonian is there, including the self Hamiltonian allows for an arbitrary duration for the measurement. Of course,  the 'measurement problem' would persist in either case even though the system-apparatus correlations have been established as the 'off-diagonal' elements (Eq.(23)) are non-vanishing in both position and momentum representations, as expected for unitary evolution. One would need to appeal to decoherence through environmental interaction for the disappearance of these elements. However, as stated before, our focus here is just the nature of the system-apparatus entangled state formed during a measurement-like interaction under pure unitary evolution and the correlations they contain. In the  case analyzed here, the  initial state of the apparatus is a momentum eigenstate and our results clearly show that system-apparatus correlations are established in the momentum basis and, surprisingly, do not seem to depend on whether or not the self Hamiltonian is included.

\subsection{Apparatus in initial position eigenstate \label{S3}}
Now, let us consider the case where the apparatus is in a position eigenstate. Though eigenstates of position are delta functions in space which are idealizations and not expected to exist in Nature, let us consider an ideal situation where the apparatus is in such a state:
\begin{equation}
\psi(z)=\delta(z-z_{0}).
\end{equation}
The  initial density matrix in the partial Fourier transform representation will be:
\begin{equation}
\rho(Q, r, 0)=e^{iz_{0}Q} \delta(r).
\end{equation} 
It can easily be checked that the solutions in the partial FT representation are:
\begin{eqnarray}
\rho_{d}(Q, r, 0)&=& \exp \{ i z_{0} Q \mp \frac{i\epsilon r t}{\hbar} \\ \nonumber
 &&\mp\frac{i \epsilon Q t^{2}}{2m} \} \delta (r + \frac{\hbar Q t}{m}). \nonumber
\end{eqnarray}
As before, if we analyze the situation where the time scale for the measurement is extremely small compared to the intrinsic time scales in the system, the self Hamiltonian can be ignored and the solutions for $\rho_d$ with just the interaction Hamiltonian (last term in eq.(7)) is:
\begin{equation}
\rho_{d}(Q, r, 0)=\exp \{ i z_{0} Q \mp \frac{i\epsilon r t}{\hbar} \} \delta (r).
\end{equation}
In both cases, the solutions can be examined in position and momentum representations: 
\begin{eqnarray}
\rho_{d}(R,r,t)&=&\exp \{\frac{i\epsilon tr}{\hbar} -\frac{mr}{\hbar t}(z_{0}+R\pm\frac{\epsilon t^{2}}{2m})\}, \\
\rho_{d}(Q, q, t)&=&\exp \{ i z_{0}Q \pm \frac{i \epsilon q t^{2}}{2m}-\frac{i \hbar Qqt}{m}\}.
\end{eqnarray}
Setting $Q=0$ and $r=0$, it is quite obvious from (30) and (31) that there are no correlations established between the spin (system) and the apparatus states either in the position representation or the momentum representation when the apparatus starts off in an initial position eigenstate. The off-diagonal elements of the density matrix in the partial FT representation is:
\begin{equation}
 \rho_{od}=\exp \{i z_{0} (Q \pm \frac{2 \epsilon t}{\hbar}) \pm \frac{2 i \lambda t}{\hbar} \} \delta (r + \frac{\hbar Q t }{m} \pm \frac{\epsilon t ^{2}}{m}),
\end{equation} 
which, as expected, does not show any decay. If the self Hamiltonian of the system is ignored, the solutions  to the diagonal density matrix in the position and momentum representations can be easily seen to be
\begin{eqnarray}
\rho_{d}(R,r,t)&=& \delta (R+ z_)}) \delta (r) e^{\pm \frac{i \epsilon r t}{\hbar} \\
\rho_{d}(Q, q, t)&=& e^{i z_{0}Q}.
\end{eqnarray}
It is obvious that in this case again, there are no correlations established between the system and apparatus states either in the postion or the momentum basis. The off-diagonal density matrix
\begin{equation}
 \rho_{od}=\exp \{ i z_{0} (Q \pm \frac{2 \epsilon t}{\hbar})
 \end{equation}
 show no decay, as before. 
\subsection{Apparatus as an initial Gaussian wave packet}
Finally, we examine the situation when the apparatus is in an initial state of a Gaussian wavepacket:
\begin{equation}
\psi (z,0)=\frac{1}{\sqrt{\sigma \sqrt{\pi}}} \exp \{ \frac{-z^{2}}{2 \sigma^{2}} \}.
\end{equation}
Here $\sigma$ is the width of the wavepacket. The density matrix corresponding to this initial state in the partial Fourier Transform representation is:
\begin{equation}
\rho(Q, r,0)=\exp \{\frac{-Q^{2} \sigma^{2}}{4} - \frac{-r^{2}}{4 \sigma^{2}} \}.
\end{equation}
It is is easy to see that the solution for the diagonal elements of the density matrix (in the partial FT representation) is:
\begin{eqnarray}
\rho_{d}(Q, r,t)&=&\exp \{\frac{-Q^{2} \sigma^{2}}{4} - \frac{\hbar^{2} Q^{2}t^{2}}{4m^{2}}-\frac{\hbar Q tr}{2 \sigma^{2}m} \\ \nonumber
&& -\frac{r^{2}}{4\sigma^{2}} \mp \frac{i \epsilon Q t^{2}}{2m} \mp \frac{i \epsilon r t}{\hbar}\}.
\end{eqnarray}
As in the previous two cases, if we analyze the situation where the time scale for the measurement is extremely small compared to the intrinsic time scales in the system, the self Hamiltonian can be ignored and the solutions for $\rho_d$ with just the interaction Hamiltonian (last term in Eqn. (7)) is
\begin{equation}
\rho_{d}(Q, r,t)=\exp \{\frac{-Q^{2} \sigma^{2}}{4}-\frac{r^{2}}{4 \sigma^{2}} \pm \frac{i \epsilon r t}{\hbar} \}.
\end{equation}
In the following we examine the solutions in the position and momentum representations for both cases.
\subsubsection{With self Hamiltonian}
It can easily be checked that in the case when the self Hamiltonian is included, the solutions in the position representation are:
\begin{eqnarray}
\rho_{d}(R, r, t)&=&\frac{2}{\sigma}\sqrt{\frac{\pi}{N(t)}}\exp \{-\frac{r^{2}}{4 \sigma^{2}}\mp\frac{i \epsilon t r}{\hbar}- \\ \nonumber
&&\frac{4}{\sigma\sqrt{N(t)}} \Big ( R \mp \frac{\epsilon t^{2}}{2m}-\frac{i \hbar r t}{2 \sigma^{2} m}\Big )^{2} \},
\end{eqnarray}
where
\begin{equation}
N(t)=1+\frac{\hbar^{2} t^{2}}{m^{2}\sigma^{2}}.
\end{equation}
On setting $r=0, R=z$, clearly (40) contains system-apparatus correlations. The up and down spins are correlated with the apparatus states corresponding to
\begin{equation}
\rho_{d}(z,t)=\frac{2}{\sigma}\sqrt{\frac{\pi}{N(t)}}\exp \{-\frac{4}{\sigma\sqrt{N(t)}} \Big ( z \mp \frac{\epsilon t^{2}}{2m}\Big )^{2} \}.
\end{equation}
In the momentum representation, it can be seen that the solution is 
\begin{eqnarray}
\rho_{d}(Q, q, t)&=& 2 \sigma \sqrt {\pi} \exp \{ - 4\sigma^{2} \Big( q \mp \frac{\epsilon t}{\hbar} + \frac{i \hbar Q t}{2 \sigma^{2} m} \Big)^{2} \\ \nonumber
&& -\frac{\sigma^{2} Q^{2} N(t)}{4} \mp \frac{i \epsilon Q t^{2}}{2m} \}. \\ \nonumber
\end{eqnarray}
Setting $Q=0, q=p$ gives the diagonal elements in momentum representation. It can easily be seen that these are
\begin{equation}
\rho_{d}(p, t))=2 \sigma \sqrt {\pi} \exp \{ - 4\sigma^{2} \Big( p \mp \frac{\epsilon t}{\hbar} \Big)^{2} \}.
\end{equation}
Once again, one can see that there are clear system-apparatus correlations in the momentum basis as well. From (42) and (44) it is evident that the spin-up and spin-down states are correlated with Gaussians in both position and momentum space and the 'separation' between the peaks (corresponding to spin-up and spin-down) in each case is:
\begin{eqnarray}
\Delta z &=&\frac{\epsilon t^{2}}{m}, \\
\Delta p &=& \frac{2 \epsilon t}{\hbar}.
\end{eqnarray}
The pure entangled state of the system and the apparatus is akin to a 'cat state'. It contains one-to-one correlations between the system and apparatus states, thus containing 'pointer states'. As the dynamics is purely unitary and there is no dissipation/decoherence involved, one would expect that the establishment of these correlations are not enough to effect a measurement as the off diagonal elements of the density matrix have not vanished and are still there. The off diagonal elements of the density matrix in the partial Fourier transform representation is:
\begin{eqnarray}
\rho_{od}(Q,r,t)&=&\exp \{ \pm \frac{2i\lambda t}{\hbar} - \frac{\sigma^{2}}{4}(Q\pm \frac{2\epsilon t}{\hbar})^{2}  \nonumber \\
&& -\frac{1}{4 \sigma^{2}} \Big ( r +\frac{\hbar Q t}{m} \pm \frac{\epsilon t^{2}}{m} \Big )^{2}-\frac{\sigma^{2} Q^{2}}{4}  \nonumber \\
&& -\frac{\sigma^{2} \epsilon Q t}{\hbar} -\frac{r^{2}}{4 \sigma^{2}} - \frac{\hbar^{2} Q^{2}t^{2}}{4 \sigma^{2} m^{2}} \nonumber \\
&&-\frac{r}{2 \sigma^{2}} \Big ( \frac{\hbar Q t}{m} \pm \frac{\epsilon t^{2}}{m} \Big ) - \frac{\hbar^{2} Q t^{3}}{4 \sigma^{2} m^{2}} \} \nonumber \\
&& \otimes \exp \{-\frac{\epsilon^{2} t^{4}}{4 \sigma^{2} m^{2}}-\frac {\sigma^{2} \epsilon^{2} t^{2}}{\hbar^{2}} \}.
\end{eqnarray}
From (45) and (46) one can see that the last term in the above expression corresponds to
\begin{equation}
\exp \{-\frac{\Delta z^{2}}{4 \sigma^{2}}-\frac{\sigma^{2} \Delta p^{2}}{4} \}.
\end{equation}
(47) shows the presence of 'decay ' terms in the expression for the off diagonal elements. These terms are proportional to the square of the 'separations', $\Delta z$ and $\Delta p$ between the pointer states, both of which increase with time (see Eqs. 45 and 46). Thus, even in the absence of any decoherence mechanism, the off diagonal elements of the density matrix are diminished. This obviously identifies localized Gaussian wavepackets as the preferred choice for the initial state of the apparatus in the sense of both establishing a one-to-one correlation in the diagonal spin elements and leading to a diminishing of the off diagonal elements. 
\subsubsection{Without  self Hamiltonian}
Finally we look at the solutions for the case where the self Hamiltonian is ignored. The diagonal and off-diagonal elements for the spin in the partial Fourier transform representation are:
\begin{eqnarray}
\rho_{d}(Q,r,t) &=& \exp \{-\frac{Q^{2}\sigma^{2}}{4} - \frac{r^{2}}{4 \sigma^{2}} \pm \frac{i \epsilon r t}{\hbar}\}\\
\rho_{od}(Q, r, t) &=& \exp \{ \pm \frac{2i\lambda t}{\hbar} -\frac{Q^{2} \sigma^{2}}{4}-\frac{r^{2}}{4\sigma^{2}}\\ \nonumber
&& - \frac{4 \epsilon t Q \sigma^{2}}{\hbar} \} \otimes \exp \{-\frac {\sigma^{2} \epsilon^{2} t^{2}}{\hbar^{2}} \}.
\end{eqnarray}
It can be easily checked from (49) that the diagonal elements in the position and momentum representations are
\begin{eqnarray}
\rho_{d}(z,t)&=&\frac{2 \sqrt{\pi}}{\sigma} \exp \{-\frac{4 z^{2}}{\sigma^{2}} \}, \\
\rho_{d}(p,t) &=& 2 \sigma \sqrt{\pi} \exp \{ -4 \sigma^{2} \Big ( p \mp \frac{\epsilon t}{\hbar} \Big )^{2} \}.
\end{eqnarray}
Surprisingly, (51) clearly indicates that there are no system-apparatus correlations established in the position basis.  By contrast, (52) shows that the correlations with the system states are in the  momentum basis and are exactly of the same form as in the case where the self Hamiltonian was included (see Eq.44). Correspondingly, the off diagonal elements (52) contain the same decay term
\begin{equation}
\exp \{-\frac {\sigma^{2} \epsilon^{2} t^{2}}{\hbar^{2}} \} =\exp \{-\frac{\sigma^{2} \Delta p^{2}}{4} \}. 
\end{equation}
Clearly, the off-diagonal terms diminish,  but this time the rate at which they do so depends only on $\Delta p^{2}$ and not on both $\Delta p^{2}$ and $\Delta z^{2}$ as in the case where the self Hamiltonian was included (see eqs.(47) and (48)).
\section{Conclusions}
We have analyzed the simple situation of a spin-1/2 particle coupled to its position degree of freedom leading to the formation of a 'system-apparatus' entangled state - a precursor to a measurement-like scenario. This model was analyzed for three different initial states of the apparatus: a momentum eigenstate, a position eigenstate, and a localized Gaussian wave packet. For each initial state, the solutions were analyzed for the case when the  self Hamiltonian was included and for the case when it was ignored. Our analysis leads to the following conclusions: (a) even in the absence of any environment-induced decoherence there is clearly a 'preferred state' of the apparatus in terms of the establishment of measurementlike one-to-one correlations in the pure system-apparatus entangled state. Of the three initial states, the Gaussian wave packet for the apparatus in clearly preferred; (b) When the initial state of the apparatus is a momentum eigenstate or a position eigenstate, the system-apparatus correlations are the same as those when the self Hamiltonian is included and when it is ignored. However, when the initial state of the apparatus is a Gaussian wavepacket, these two cases give different results - while pointer states emerge in both $z$ and $p$ basis with the full Hamiltonian, they emerge only in the $p$ baisis for the case where the self Hamiltonian is ignored. This simple analysis, which involves only unitary Schr\"{o}dinger dynamics significantly illustrates elements of quantum measurement like the nature of the correlations established in the system-apparatus entangled state and illuminates that even in this simple situation a 'preferred' candidate for the apparatus state exists.
\section{Acknowledgments}
The author wishes to thank the Centre for Philosophy and Foundations of Science, New Delhi and acknowledges financial support from the Department of Science and Technology, Government of India.

\end{document}